\documentclass[12pt,preprint]{aastex}
 
\shorttitle{Gas Absorption by Edge-on HD32297 Disk}
\shortauthors{Redfield}
 
\begin{document}
 
\newcommand{\php}[0]{\phantom{--}}
\newcommand{\kms}[0]{km~s$^{-1}$}
 
\title{Gas Absorption Detected from the Edge-on Debris Disk Surrounding HD32297}
 
\author{Seth Redfield\altaffilmark{1}}
\affil{Department of Astronomy and McDonald Observatory, University of Texas, Austin, TX, 78712}
\altaffiltext{2}{Hubble Fellow.}
\email{sredfield@astro.as.utexas.edu}

\begin{abstract}

Near-infrared and optical imaging of HD32297 indicate that it has an
edge-on debris disk, similar to $\beta$ Pic.  I present high
resolution optical spectra of the \ion{Na}{1} doublet toward HD32297
and stars in close angular proximity.  A circumstellar absorption
component is clearly observed toward HD32297 at the stellar radial
velocity, which is not observed toward any of its neighbors, including
the nearest only 0$\farcm$9 away.  An interstellar component is
detected in all stars $>$90\,pc, including HD32297, likely due to the
interstellar material at the boundary of the Local Bubble.  Radial
velocity measurements of the nearest neighbors, BD+07\,777s and
BD+07\,778, indicate that they are unlikely to be physically
associated with HD32297.  The measured circumstellar column density
around HD32997, $\log N_{\rm NaI} \sim 11.4$, is the strongest
\ion{Na}{1} absorption measured toward any nearby main sequence debris
disk, even the prototypical edge-on debris disk, $\beta$ Pic.
Assuming that the morphology and abundances of the gas component
around HD32297 are similar to $\beta$ Pic, I estimate an upper limit
to the gas mass in the circumstellar disk surrounding HD32297 of
$\sim$0.3 $M_{\oplus}$.

\end{abstract}
 
 
\section{Introduction}

Debris disk systems provide a look at an intermediate stage of stellar
system evolution.  They represent the transition between the early
formation of stars and planets in a primordial protoplanetary disk as
seen toward pre-main sequence stars, and the mature stage of an
evolved system, like our solar system, which is clear of all
primordial material and retains only a hint of secondary products
(e.g., zodiacal dust), the final remnants of the stellar and planetary
formation process.  Although a debris disk has lost most of its
primordial material, the observed infrared luminosity of circumstellar
dust, caused by collisions of planetismals and other small bodies, is
typically several orders of magnitude larger than estimated for the
Kuiper and asteroid belts in our solar system \citep{backman93}.  Ever
since the detection of dusty circumstellar material around main
sequence stars via infrared excesses \citep{aumann84}, researchers
have been looking for circumstellar gas phase absorption
\citep{hobbs85}.  Of the initial major infrared excess main sequence
stars, only $\beta$ Pic showed gas phase absorption in optical
absorption lines (e.g., \ion{Ca}{2} and \ion{Na}{1}), due to its disk
morphology and edge-on orientation \citep{smith84}.  Such on
orientation provides a unique opportunity to simultaneously measure
both the dust and gas components of a debris disk, at an interesting
transition near the end of stellar and planetary formation.

Only a few other edge-on debris disks have been found since, including
$\beta$ Car \citep{lagrangehenri90}, HD85905 \citep{welsh98}, HR10
\citep{lagrangehenri90hr10}, and AU Mic (\citeauthor*{kalas04}
\citeyear{kalas04}; \citeauthor{roberge05} \citeyear{roberge05}).
\citet*{redfield07lismdd} observed $\beta$ Car, HD85905, HR10 with the
{\it Spitzer Space Telescope} and did not find strong infrared
excesses toward any of them, although an optical monitoring campaign
showed clear signs of gas variability, as noted by researchers
earlier.  However, the magnitude of circumstellar absorption in these
systems is lower than observed toward $\beta$ Pic.

Long \ion{Ca}{2} monitoring campaigns of $\beta$~Pic
\citep[e.g.,][]{petterson99}, find significant short-term absorption
variability.  This variability can be explained by gas clouds very
close to the star, which are caused by evaporating, star-grazing,
km-sized objects, simply referred to as, Falling Evaporating Bodies
\citep[FEB's;][]{beust94}.  A strong ``stable'' component, at rest in
the stellar reference frame, is also detected toward $\beta$~Pic
\citep[e.g.,][]{crawford94}.  The distribution of gas in this
component, contrary to the variable component located very close to
the star, is dispersed throughout the extended dust disk
\citep{brandeker04}.

A ``stable'' absorption component in a gas phase resonance line can be
caused by either intervening circumstellar or interstellar gas.
Measuring the interstellar medium (ISM) along the line of sight and in
the locality surrounding a circumstellar disk candidate, is critical
to characterizing any ``contaminating'' ISM absorption
\citep{crawford01,redfield07lismdd}.  In particular, the Sun resides
in a large scale ISM structure known as the Local Bubble, whose boundary
at $\sim$100\,pc is defined by a significant quantity of interstellar
material \citep{lallement03}.  If a ``stable'' absorption component is
observed at the stellar radial velocity, and similar absorption is not
detected toward any proximate stars, it is likely that the absorption
component is caused by circumstellar material.

Using near-infrared scattered light observations taken with the {\it
Hubble Space Telescope}, \citet*{schneider05} discovered that the
debris disk surrounding HD32297 has an edge-on orientation.  Disk
emission extends out to $\sim$400\,AU in their observations, while
radii $<$33.6\,AU are occulted by the coronagraphic obstacle.  Optical
scattered light observations by \citet{kalas05} confirmed this
orientation and extended the range of disk emission to $\sim$1680\,AU.
The edge-on orientation of HD32297 makes it an ideal target for gas
phase absorption measurements.

\section{Observations}

Observations of the \ion{Na}{1} D doublet (5895.9242 and
5889.9510\,\AA) toward HD32297 were made over several epochs.  The
\ion{Na}{1} doublet is among the strongest transitions in the optical
wavelength band, appropriate for observing interstellar
\citep{redfield06} and circumstellar \citep{lagrangehenri90}
absorption toward nearby stars.  In addition, several stars in close
angular proximity to HD32297 were observed, in order to reconstruct
the ISM absorption profile along the line of sight.  Stellar
parameters of the observed targets are given in
Table~\ref{tab:basics}, and the observational parameters are listed in
Table~\ref{tab:fits}.

High resolution optical spectra were obtained using the Coud\'{e}
Spectrometer on the 2.7m Harlan J. Smith Telescope at McDonald
Observatory.  The spectra were obtained at a resolution of
$R\,\equiv\,\lambda/\Delta\lambda\,\sim\,$240,000, using the
2dcoud\'{e} Spectrograph \citep{tull95} in the CS21 configuration.
The data were reduced using Image Reduction and Analysis Facility
\citep[IRAF;][]{tody93} and Interactive Data Language (IDL) routines
to subtract the bias, flat field the images, remove scattered light
and cosmic ray contamination, extract the echelle orders, calibrate
the wavelength solution, and convert to heliocentric velocities.
Wavelength calibration images were taken using a Th-Ar hollow cathode
before and after each target.

Numerous weak water vapor lines are commonly present in spectra around
the \ion{Na}{1} doublet, and must be modeled and removed, in order to
measure an accurate interstellar (or circumstellar) \ion{Na}{1}
absorption profile.  I use a forward modeling technique demonstrated
by \citet{lallement93} to remove telluric line contamination in the
vicinity of the \ion{Na}{1} D lines, with a terrestrial atmosphere
model (AT - Atmospheric Transmission program, from Airhead Software,
Boulder, CO) developed by Erich Grossman.  With two \ion{Na}{1}
absorption lines, it is straightforward to identify contaminating
telluric absorption.

All absorption lines were fit using standard methods \citep[e.g.,
\S2.2 in][]{redfield04sw}.  Gaussian absorption components are fit to
both \ion{Na}{1} D lines simultaneously using atomic data from
\citet{morton03}, and then convolved with the instrumental line spread
function.  Fitting the lines simultaneously reduces the influence of
systematic errors, such as continuum placement and contamination by
weak telluric features.  The free parameters are the central velocity
($v$), the line width or Doppler parameter ($b$), and the column
density ($N$) of \ion{Na}{1} ions along the line of sight.  The fits
are shown in Figure~\ref{fig:hd32297_figna1} and fit parameters with
1$\sigma$ statistical errors are listed in Table~\ref{tab:fits}.

In addition, the spectra were used to estimate the stellar radial
velocity ($v_{\rm R}$) and projected stellar rotation ($v\sin i$) for
HD32297, BD+07 777s, and BD+07 778 (see Table~\ref{tab:basics}),
quantities not listed in SIMBAD for these targets.  The radial
velocities of all 3 objects differ significantly, and therefore it is
unlikely that they are physically associated.  Note that the radial
velocity of HD32297 ($v_{\rm R} \sim +20$ km~s$^{-1}$) is measured
from broad \ion{Na}{1} and H$\alpha$ stellar absorption
lines, and therefore is not tightly constrained.

\section{Identification of Circumstellar Absorption Toward HD32297}

\begin{figure}
\epsscale{.75}
\plotone{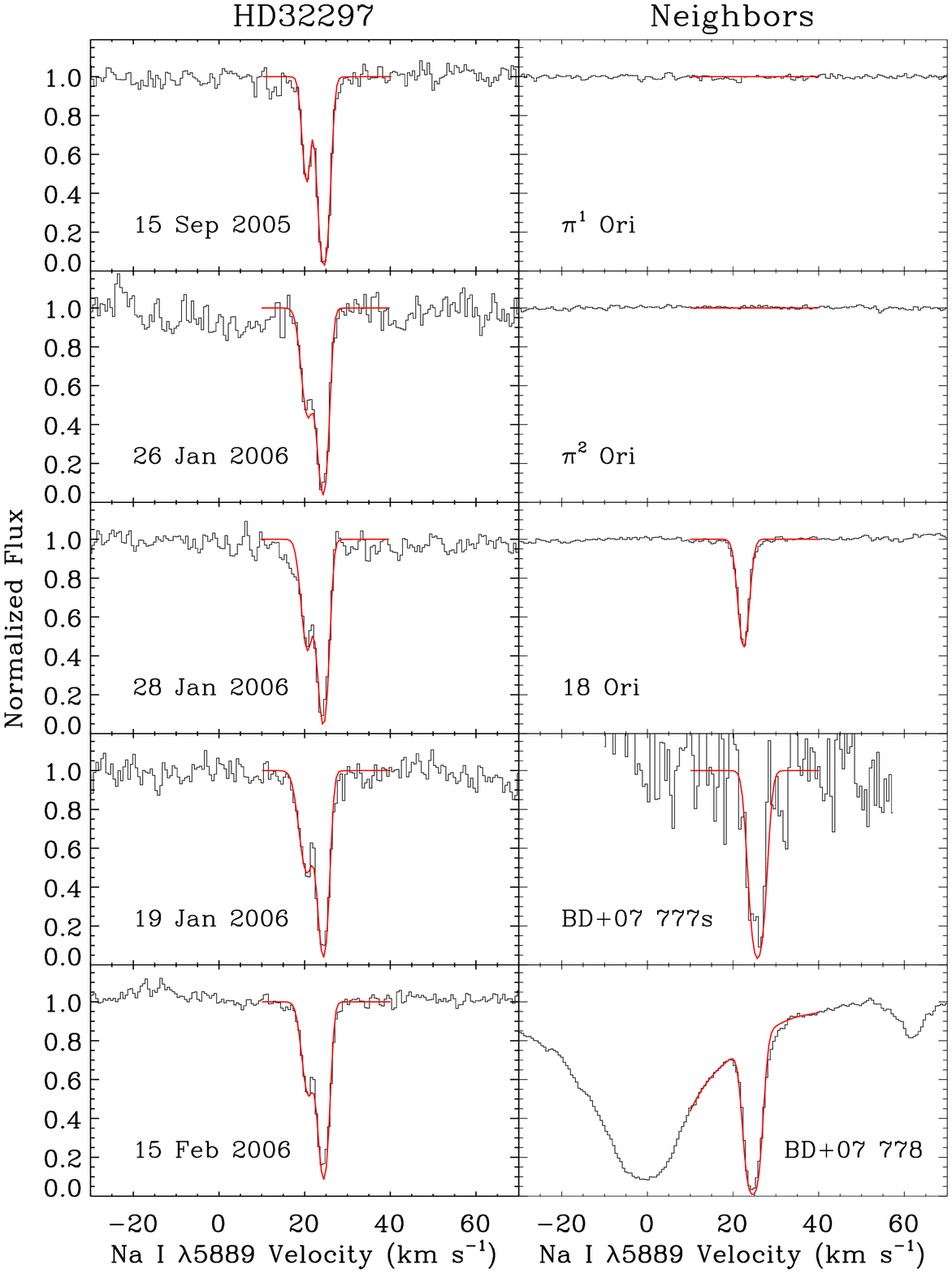}
\caption{\ion{Na}{1} absorption lines toward HD32297 ({\it left}) and
toward stars proximate to HD32297 ({\it right}). All spectra are flux
normalized such that stellar features are removed, except for BD+07
778, where the stellar feature is retained.  All objects beyond
60\,pc, including HD32297, show a significant interstellar component
at $\sim$24.5 km~s$^{-1}$, whereas only HD32297 has an additional
absorption feature $\sim$20.5 km~s$^{-1}$, presumably circumstellar.
Temporal variability can be seen in the circumstellar component and
redward in HD32297 over several observational epochs. The absorption
in both lines of the \ion{Na}{1} doublet is fit simultaneously.  The
best fit, convolved with the line spread function, is overplotted
(red). \label{fig:hd32297_figna1}}
\end{figure}

The left column of Figure~\ref{fig:hd32297_figna1} shows that
\ion{Na}{1} absorption is clearly detected toward HD32297 in 5
observations over 5 months.  Two components are easily distinguished,
a strong component at $\sim$24.5 km~s$^{-1}$ and a weaker component at
$\sim$20.5 km~s$^{-1}$.  The \ion{Na}{1} spectral region for 5 stars
in close angular proximity to HD32297 is also shown in
Figure~\ref{fig:hd32297_figna1}.  Only a single ISM component, at
$\sim$24.2 km~s$^{-1}$, is detected in the 3 distant neighbors,
indicating that large scale interstellar material is located at a
distance between 59.4--112\,pc.  All targets located beyond this
material, including HD32297, should have a similar ISM absorption
feature.  This strong ISM absorption is probably associated with the
boundary material of the Local Bubble, which is estimated to be
$\sim$90 pc in this direction \citep{lallement03}.  If located at this
distance, the physical separation of the interstellar material
observed toward HD32297 and the material toward BD+07 777s
($\Delta\theta =$ 0$\farcm$9) is 0.025\,pc, BD+07 778 (2$\farcm$4) is
0.064\,pc, and 18 Ori (5$\fdg$1) is 8.1\,pc.  Toward HD32297's two
closest neighbors, the ISM absorption is almost identical in projected
velocity and column density to the strong absorption seen toward
HD32297, while toward 18 Ori, the absorption differs slightly in both
$v$ and $N$, indicating that any small scale morphological variations
in the Local Bubble shell are on scales $>$0.1\,pc but $<$8\,pc.
Small scale variations in the Local Bubble shell have been detected by
\citet*{redfield06sins} on scales $\sim$0.5\,pc.

It is unlikely that the unique 20.5 km~s$^{-1}$ feature observed
toward HD32297 is caused by a small scale interstellar structure.
Although small ISM structures (0.01--2.0\,pc) have been observed
\citep*[e.g.,][]{ferlet85,meyer96}, it is more likely that the unique
feature is due to absorption in the circumstellar environment
surrounding HD32297 because (1) HD32297 is known to be an edge-on
debris disk, (2) no similar absorption is detected in the very close
neighboring sightlines (0.03--0.08\,pc), and (3) the absorption
matches the stellar radial velocity.

\section{Temporal Variability of Circumstellar Component}

\begin{figure}
\epsscale{.75}
\plotone{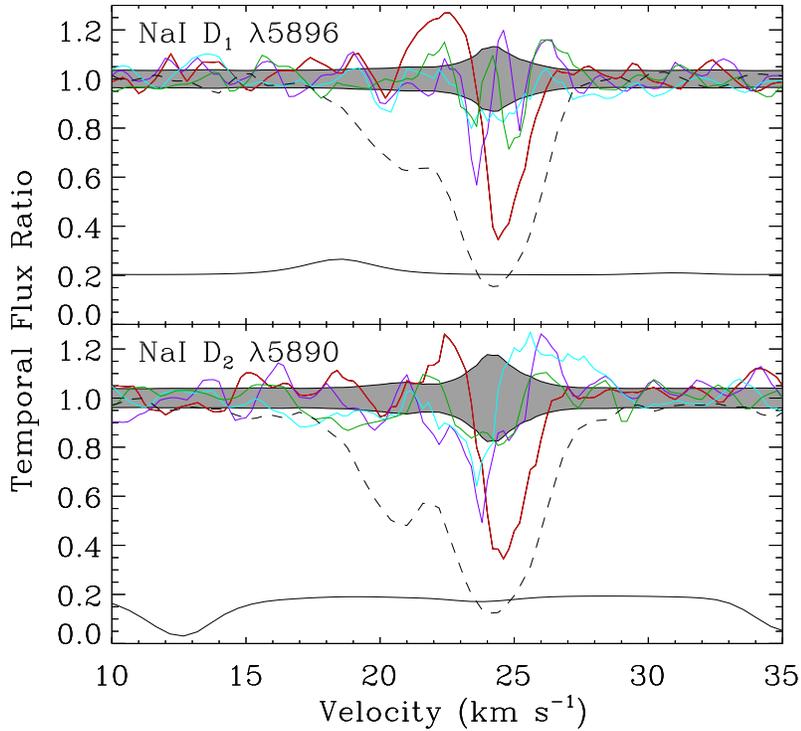}
\caption{Difference spectra of all HD32297 observations for both
\ion{Na}{1} lines, relative to the final spectrum taken 2006 Feb 15,
which is plotted as the dashed line.  The gray shaded region indicates
the ratio error band.  The ratio of the telluric water vapor spectra
of 2005 Sep 15 and 2006 Feb 15 is shown offset toward the bottom of
each plot.  The most notable temporal variation occurs between the
2005 Sep 15 observation (red), where the flux variation, centered at
$\sim$22.5 km~s$^{-1}$, is $\sim$5$\sigma$.  This redshifted
variability is common in gas absorption toward edge-on debris disks,
although in this particular instance the variability is partially
masked by the strong ISM component. \label{fig:hd32297_diff}}
\end{figure}

Temporal variability is also a hallmark of circumstellar material
(e.g., \citealt*{ferlet87}, \citealt{petterson99},
\citealt{redfield07lismdd}).  To search for variability,
Figure~\ref{fig:hd32297_diff} shows difference spectra of all
observations.  Some indication of temporal variability on time scales
of months is detected.  For example, between the 2005 Sep and 2006 Feb
observations, the $\sim$20.5 km~s$^{-1}$ feature became stronger and
the separation between the circumstellar and interstellar features
became less distinct, despite the fact that the 2006 Feb observations
were made at a slightly higher resolving power.  The redshifted
variability seen between 2005 Sep and 2006 Feb is $\sim$5$\sigma$
above the standard deviation.  The same pattern is seen in both
\ion{Na}{1} lines, indicating that the telluric contamination is not
causing the variation.  Slight changes in the resolving power of our
instrument could mimic this variable behavior, differentially moving
light from the cores of the line to the wings (or vice versa).
However, resolution variability should cause (1) symmetric features in
the wings of the line whereas we see a feature only to the blue of the
ISM feature and not to the red, and (2) should have a stronger effect
on stronger absorption features, whereas the feature is roughly
identical in both lines, which could be caused if the absorbing
material covers only a fraction of the stellar disk, as has been seen
toward $\beta$ Pic \citep{vidalmadjar94}.

This data alone provides only a subtle indication of temporal
variation in \ion{Na}{1}, partially because any significant absorption
toward the red, is masked by the strong ISM feature.  Redshifted
circumstellar absorption dominates the \ion{Ca}{2} gas absorption
variability toward $\beta$ Pic \citep[e.g.,][]{petterson99}, while no
temporal variability has ever been detected in \ion{Na}{1} toward
$\beta$ Pic, only the ``stable'' absorption component is seen in this
ion \citep{welsh97}.  Circumstellar variability in \ion{Na}{1} has
been detected in other edge-on debris disks, e.g., $\beta$ Car,
HD85905, and HR10 \citep{welsh98,redfield07lismdd}.  Any redshifted
absorption occuring in this object could cause fluctuations in the
measured column density of the ``constant'' ISM feature.  Little
evidence for variability is found toward the blue.

\section{Gas Disk Mass}

These observations indicate that HD32297 has the strongest \ion{Na}{1}
circumstellar disk signature detected around a nearby main sequence
debris disk star.  Even compared to $\beta$ Pic, the prototypical
edge-on debris disk with \ion{Na}{1} absorption column densities of
$\log N_{\rm NaI} \sim 10.69$--$10.73$ \citep{hobbs85,welsh97}, the
gas disk around HD32297, with $\log N_{\rm NaI} \sim 11.4$, has
5$\times$ the \ion{Na}{1} column density.  A crude estimate of the gas
mass surrounding HD32297 can be made if it is assumed to have the same
morphology and abundances as the stable gas around $\beta$ Pic.
Although the observations of HD32297 indicate some red-shifted
temporal variability, much of the gas is stable over all observations.
Using $\beta$ Pic as a proxy, the variable gas is likely located very
close to the star \citep*{lagrange00}, while the stable gas at rest in
the stellar frame, likely traces the bulk dust disk
\citep{brandeker04}.  For this calculation, I assume all the gas is in
the stable component, and therefore this gas mass estimate should be
considered an upper limit.  The morphology of the disk is assumed to
follow a broken power law density profile, as fit to the \ion{Na}{1}
emission profile of the $\beta$ Pic disk (see Equation~1 of
\citealt{brandeker04}), and assumed to extend out to the edge of the
debris disk at $\sim$1680\,AU \citep{kalas05}.  The abundances in the
HD32297 disk are assumed to be similar to $\beta$ Pic
\citep{roberge06}, where the ratio $N($\ion{H}{1}$)/N($\ion{Na}{1}$)
\lesssim 8.8 \times 10^8$, is based on $\beta$ Pic \ion{Na}{1}
measurements by \citet{brandeker04} and \ion{H}{1} limits by
\citet{freudling95}.  Given these assumption, I calculate a gas mass,
distributed through the bulk debris disk surrounding HD32297 at
$M_{\rm gas} \sim 0.3 M_{\oplus}$.

Future observations are planned to continue monitoring the temporal
variability of the circumstellar gas toward HD32297 to determine the
ratio of stable to variable gas, and measure the \ion{Ca}{2} gas disk
absorption, in order to independently measure the \ion{Ca}{2} to
\ion{Na}{1} ratio.  A more definitive detection of temporal
variability may require monitoring excited lines which will show
circumstellar absorption, but not the strong interstellar feature.

\section{Conclusions}

I present the first high resolution optical spectra of the \ion{Na}{1}
doublet toward the debris disk HD32297 and stars in close angular
proximity.  A summary of results include:

(1) Two absorption components are detected toward HD32297, while only
one is detected toward its proximate neighbors located at a
comparable distance.  The extra absorption component in the spectrum
of HD32297, which is also at rest in the stellar reference frame, is
therefore likely caused by circumstellar material.\\
\indent(2) The ISM absorption is similar among HD32297 and its two
closest neighbors, and is likely due to absorption from the shell that
defines the boundary of the Local Bubble.  Some variation in Local
Bubble absorption is detected toward 18 Ori.\\
\indent(3) Radial velocities of HD32297, BD+07 777s, and BD+07 778 are
measured and differ significantly, indicating that they are likely not
physically associated.\\
\indent(4) Some indication of temporal variability is detected over several
epochs of observations.  Instrumental resolution variations and
masking by the strong ISM absorption, make a definitive detection of
circumstellar \ion{Na}{1} variability difficult.\\
\indent(5) The measured circumstellar feature toward HD32297 ($\log N_{\rm
NaI} \sim 11.4$) is the strongest such absorption measured toward any
nearby main sequence debris disk, $\sim$5 times greater than the
column density of the prototypical edge-on debris disk, $\beta$ Pic.\\
\indent(6) If the morphology and abundances of the stable gas component
around HD32297 are assumed to be similar to $\beta$ Pic, I estimate an
upper limit to the gas mass in the circumstellar disk surrounding
HD32297 of $\sim$0.3 $M_{\oplus}$.

\acknowledgements 

Support for this work was provided by NASA through Hubble Fellowship
grant HST-HF-01190.01 awarded by the Space Telescope Science
Institute, which is operated by the Association of Universities for
Research in Astronomy, Inc., for NASA, under contract NAS 5-26555.  I
would like to thank D.\ Doss, G.\ Harper, and A.\ Brown, for their
assistance with these observations.  The insightful comments by the
anonymous referee were very helpful.

{\it Facilities:} \facility{Smith (CS21)}



\clearpage

\clearpage

\clearpage

\begin{deluxetable}{lllcllcclcc}
\tablewidth{0pt}
\tabletypesize{\scriptsize}
\tablecaption{Stellar Parameters for HD32297 and Nearby Stars\tablenotemark{a}\label{tab:basics}}
\tablehead{ HD & Other & Spectral & $m_{\rm V}$ & $v_{\rm R}$ & $v \sin i$ & $l$ & $b$ & Distance\tablenotemark{b} & $\Delta\theta$ & $\Delta r_{\rm pos}$\tablenotemark{c} \\
\# & Name & Type & (mag) & (km s$^{-1}$) & (km s$^{-1}$) & (deg) & (deg) & (pc) & (deg) & (pc) }
\startdata
32297  & BD+07 777    & A0       & 8.13 & $\sim$+20\tablenotemark{d} & $\sim$80\tablenotemark{d} & 192.83 & --20.17 & 112$^{+15}_{-12}$ & 0.0000 & 0.0 \\
\hline
\nodata & BD+07 777s   & G0\tablenotemark{e} & 10.2 & --55\tablenotemark{d} & $\sim$2\tablenotemark{d} & 192.85 & --20.17 & \nodata           & 0.0156 & 0.030 \\
32304  & BD+07 778    & G5       & 6.87 & --1.4\tablenotemark{d} & $\sim$3.5\tablenotemark{d} & 192.88 & --20.17 & 134$^{+19}_{-15}$ & 0.0406 & 0.079 \\
30739  & $\pi^2$ Ori  & A1Vn     & 4.35 & +24    & 212 & 189.82 & --21.83 & 59.4$^{+4.9}_{-4.2}$ & 3.2653 & 3.4 \\
31295  & $\pi^1$ Ori  & A0V      & 4.66 & +13    & 120 & 189.35 & --20.25 & 37.0$^{+1.3}_{-1.2}$ & 3.2748 & 2.1 \\
34203  & 18 Ori       & A0V      & 5.52 & --8.2  & 70  & 191.29 & --15.25 & 112.9$^{+10.4}_{-8.8}$ & 5.1306 & 10.1 \\

\enddata
\tablenotetext{a}{All values from SIMBAD unless otherwise noted.}
\tablenotetext{b}{Distances calculated from {\it Hipparcos} parallaxes.}
\tablenotetext{c}{Physical separation in the plane of the sky from HD32297, at the distance of the closest partner.}
\tablenotetext{d}{Measured from spectra presented in this paper.}
\tablenotetext{e}{Spectral type based on $B-V = 0.57$.}
\end{deluxetable}

\clearpage

\begin{deluxetable}{lllcccccc}
\tablewidth{0pt}
\tabletypesize{\scriptsize}
\tablecaption{Observational and Absorption Fit Parameters for HD32297 and Nearby Stars\label{tab:fits}}
\tablehead{ HD & Other & Date & $v_{\rm atm}$\tablenotemark{a} & $\langle FWHM_{\rm ThAr} \rangle$\tablenotemark{b} & S/N\tablenotemark{c} & $v$ & $b$ & $\log N$  \\
\# & Name & & (km s$^{-1}$) & (km s$^{-1}$) & & (km s$^{-1}$) & (km s$^{-1}$) & (cm$^{-2}$)}
\startdata
32297  & BD+07 777    & 2005 Sep 15 & +28.2 & 1.565 & 31 & 20.459 $\pm$ 0.030 & 0.38 $\pm$ 0.28 & 10.97$^{+0.25}_{-0.67}$ \\
       &              &             &       &      &    & 24.546 $\pm$ 0.012 & 0.742 $\pm$ 0.063 & 12.388$^{+0.083}_{-0.103}$ \\
32297  & BD+07 777    & 2006 Jan 26 & --22.8& 1.253 & 17& 20.50 $\pm$ 0.27 & 1.23 $\pm$ 0.55 & 11.384 $\pm$ 0.077 \\
       &              &             &       &      &    & 24.30 $\pm$ 0.12 & 0.84 $\pm$ 0.21 & 12.16$^{+0.12}_{-0.16}$ \\
32297  & BD+07 777    & 2006 Jan 28 & --23.4 & 1.266 & 22& 20.48 $\pm$ 0.23 & 1.24 $\pm$ 0.42 & 11.449$^{+0.053}_{-0.060}$ \\
       &              &             &       &      &    & 24.34 $\pm$ 0.17 & 0.82 $\pm$ 0.31 & 11.98$^{+0.25}_{-0.64}$ \\
32297  & BD+07 777    & 2006 Jan 29 & --23.7 & 1.288 & 20& 20.42 $\pm$ 0.20 & 1.71 $\pm$ 0.28 & 11.454$^{+0.049}_{-0.050}$ \\
       &              &             &       &      &    & 24.424 $\pm$ 0.086 & 0.71 $\pm$ 0.23 & 12.02$^{+0.18}_{-0.32}$ \\
32297  & BD+07 777    & 2006 Feb 15 & --27.5 & 1.199 & 39& 20.51 $\pm$ 0.26 & 1.14 $\pm$ 0.41 & 11.297$^{+0.097}_{-0.120}$ \\
       &              &             &       &      &    & 24.41 $\pm$ 0.13 & 0.94 $\pm$ 0.30 & 11.97$^{+0.15}_{-0.20}$ \\
\hline
\nodata & BD+07 777s   & 2006 Feb 16 & --27.7 & 1.238 & 6& 25.55 $\pm$ 0.28 & 1.581 $\pm$ 0.078 & 12.03$^{+0.12}_{-0.18}$ \\
32304  & BD+07 778    & 2006 Feb 16 & --27.7 & 1.244 & 52& 24.583 $\pm$ 0.083 & 1.65 $\pm$ 0.15 & 12.18$^{+0.12}_{-0.16}$ \\
30739  & $\pi^2$ Ori  & 2006 Feb 16 & --28.3 & 1.213 & 133& \nodata & \nodata & $<9.8$ \\
31295  & $\pi^1$ Ori  & 2004 Oct 18 & +21.1 & 1.761 & 100& \nodata & \nodata & $<10.1$ \\
34203  & 18 Ori       & 2006 Feb 15 & --27.3 & 1.189 & 117& 22.464 $\pm$ 0.022 & 1.150 $\pm$ 0.036 & 11.403$^{+0.040}_{-0.044}$ \\
\enddata
\tablenotetext{a}{Projected velocity of the Earth's atmosphere.}
\tablenotetext{b}{Resolution based on neighboring ThAr comparison lamp spectra assuming that the ThAr lines are fully resolved.}
\tablenotetext{c}{Signal-to-noise at core of NaI absorption line.}
\end{deluxetable}

\end{document}